\documentclass[graybox]{svmult}

\usepackage{mathptmx}       
\usepackage{helvet}         
\usepackage{courier}        
\usepackage{type1cm}        
\usepackage{makeidx}         
\usepackage{graphicx}        
\usepackage{multicol}        
\usepackage[bottom]{footmisc}

\makeindex             

\begin{document}

\title*{The INTEGRAL-OMC Scientific Archive}
\author{A. Domingo, R. Guti\'errez-S\'anchez, D. R\'{\i}squez, M. D. Caballero-Garc\'{\i}a, J. M. Mas-Hesse and E. Solano}
\authorrunning{A. Domingo et al.}
\institute{A. Domingo \at CAB/LAEFF (CSIC-INTA), POB 78, 28691 Villanueva de la Ca\~nada, Madrid, Spain\\ \email{albert@laeff.inta.es}}
%
%
\maketitle

\abstract*{The Optical Monitoring Camera (OMC) on-board the INTEGRAL satellite has, as one of its
scientific goals, the observation of a large number of variable sources previously selected. After
almost 6 years of operations, OMC has monitored more than 100\,000 sources of scientific interest.
In this contribution we present the OMC Scientific Archive (\url{http://sdc.laeff.inta.es/omc/}) which
has been developed to provide the astronomical community with a quick access to the light curves
generated by this instrument. We describe the main characteristics of this archive, as well as
important aspects for the users: object types, temporal sampling of light curves and photometric
accuracy.}

\abstract{The Optical Monitoring Camera (OMC) on-board the INTEGRAL satellite has, as one of its
scientific goals, the observation of a large number of variable sources previously selected. After
almost 6 years of operations, OMC has monitored more than 100\,000 sources of scientific interest.
In this contribution we present the OMC Scientific Archive (\url{http://sdc.laeff.inta.es/omc/}) which
has been developed to provide the astronomical community with a quick access to the light curves
generated by this instrument. We describe the main characteristics of this archive, as well as
important aspects for the users: object types, temporal sampling of light curves and photometric
accuracy.}

\section{Introduction}
\label{sec:intro}
The Optical Monitoring Camera (OMC) \cite{mas2003} observes the optical emission from the prime
targets of the $\gamma$-ray instruments on-board the ESA mission INTEGRAL \cite{winkler2003}, with
the support of the JEM-X monitor in the X-ray domain. In addition to the prime targets, OMC observes
serendipitously a large amount of optically variable objects previously selected in the OMC Input
Catalogue \cite{domingo2003}. 

The OMC is based on a refractive optics with an aperture of 50~mm
focused onto a large format CCD ($1024\times2048$ pixels) working in frame transfer mode
($1024\times1024$ pixels imaging area). With a field of view of $5^\circ\times5^\circ$ is able to
monitor sources down to magnitude $V\approx17$. The operations of INTEGRAL-OMC provide a unique
photometric capability to obtain light curves of variable stars presenting periods that can not be
addressed properly from ground-based observatories.

\begin{figure}[t]
\centering
\includegraphics[width=0.7\textwidth]{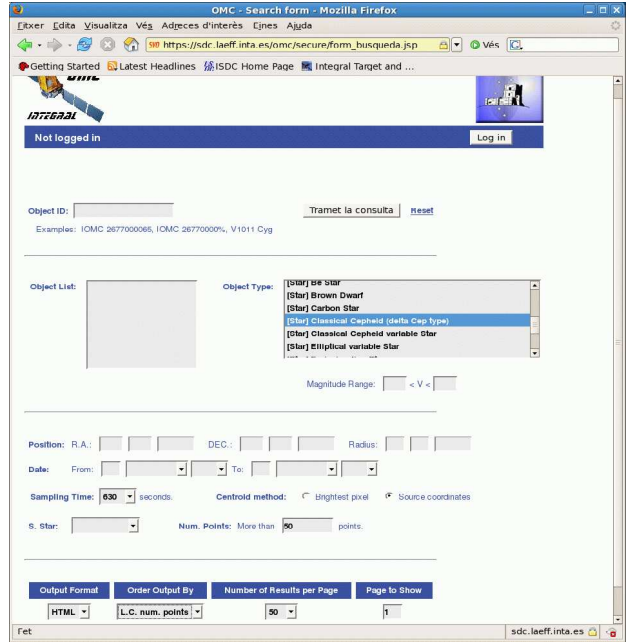}
\caption{Search capabilities available in the Archive Web Portal.}
\label{fig:portal}
\end{figure}

\section{Archive Web Portal}
\label{sec:portal}

In normal operations OMC monitors routinely around 100 sources in each field. Due to the
observational strategy of INTEGRAL, most of the sky has been already observed several times. This
allows the delivery, at the end of the mission, of a catalogue of thousands of variable sources with a
well calibrated optical magnitude, covering all kind of periods and monitored over a long interval of
time.

A scientific archive \cite{gutierrez2004} was developed to provide the astronomical community with a
quick access to the OMC light curves. This archive can be reached at
\url{http://sdc.laeff.inta.es/omc/}. It contains the data generated by the OMC, updated regularly,
and an access system capable of performing complex searches (Fig.~\ref{fig:portal}). A remarkable
point is the existence of visualization and analysis tools, available from the user's interface,
aiming at optimizing the scientific return of the OMC data.

\begin{table}[t]
\caption{OMC observed sources with known optical counterpart. Only those sources with more than 50
photometric points have been considered.}
\label{tab:sources}
\begin{tabular}{l@{\extracolsep{-1.3cm}}l@{\extracolsep{0.0cm}}r@{\extracolsep{1.1cm}}l@{\extracolsep{-1.3cm}}l@{\extracolsep{0.0cm}}r}
\hline\noalign{\smallskip}
Type & \hspace{0.5cm}Subtype & Number & Type & \hspace{0.5cm}Subtype & Number\\
\noalign{\smallskip}\svhline\noalign{\smallskip}
\multicolumn{6}{c} {\bf Variable stars}\\
\noalign{\smallskip}\hline\noalign{\smallskip}
Irregular       &                                  & 242 & Rotational      &                                  & 566\\
                & Orion                            & 118 &                 & Pulsar                           & 512\\
                & With rapid variations            &  52 &                 & a2 Canum Venaticorum             &  23\\
                & Without subtype                  &  72 &                 & RS Canum Venaticorum             &  17\\
Eruptive        &                                  & 298 &                 & Others or without subtype        &  14\\
                & Flare star                       &  72 & Pulsating       &                                  &4960\\
                & R Coronae Borealis               &  11 &                 & Mira                             &1758\\
                & T Tauri                          & 202 &                 & RR Lyrae                         &1268\\
                & Others or without subtype        &  13 &                 & Cepheid                          & 104\\
Rotational      &                                  & 566 &                 & Classical Cepheid ($\delta$ Cephei) & 265\\
                & Pulsar                           & 512 &                 & Semi-regular                     & 792\\
                & a2 Canum Venaticorum             &  23 &                 & Others or without subtype        & 773\\
Symbiotic       &                                  &  18 & Others/No type  &                                  &4294\\
\noalign{\smallskip}\svhline\noalign{\smallskip}
\multicolumn{6}{c} {\bf Composite objects}\\
\noalign{\smallskip}\hline\noalign{\smallskip}
Cataclysmic star&                                  & 352 & Eclipsing binary&                                  &1900\\
                & Nova                             & 193 &                 & Algol                            &1207\\
                & Dwarf Nova                       & 109 &                 & $\beta$ Lyrae                    & 221\\
                & Others or without subtype        &  50 &                 & W Ursae Majoris                  & 153\\
X-ray binary    &                                  & 249 &                 & Without subtype                  & 319\\
                & High Mass (HMXB)                 &  74 &                 &                                  &    \\
                & Low Mass (LMXB)                  & 162 &                 &                                  &    \\
                & Without subtype                  &  13 &                 &                                  &    \\
\noalign{\smallskip}\svhline\noalign{\smallskip}
\multicolumn{6}{c} {\bf Galaxies}\\
\noalign{\smallskip}\hline\noalign{\smallskip}
AGN             &                                  & 923 & Radio galaxy    &                                  & 196\\
                & Seyfert                          & 198 & Emission-line galaxy&                              & 146\\
                & Blazar                           &  40 & Possible Quasar &                                  & 484\\
                & Quasar                           & 628 & Others/No type  &                                  & 159\\
                & Others or without subtype        &  57 &                 &                                  &    \\
\noalign{\smallskip}\hline\noalign{\smallskip}
\end{tabular}
\end{table}

\section{Observed Sources}
\label{sec:sources}

The main scientific objective of OMC is to monitor the optical emission of the high energy sources
observed by the $\gamma$-ray and X-ray instruments, even if the optical counterpart is unknown.
However in this section we want to focus on the large amount of optically variable objects which are
observed serendipitously. In Table~\ref{tab:sources} we show a summary of these serendipitous
sources available from our Web Portal after almost 6 years of operations.

\begin{figure}[t]
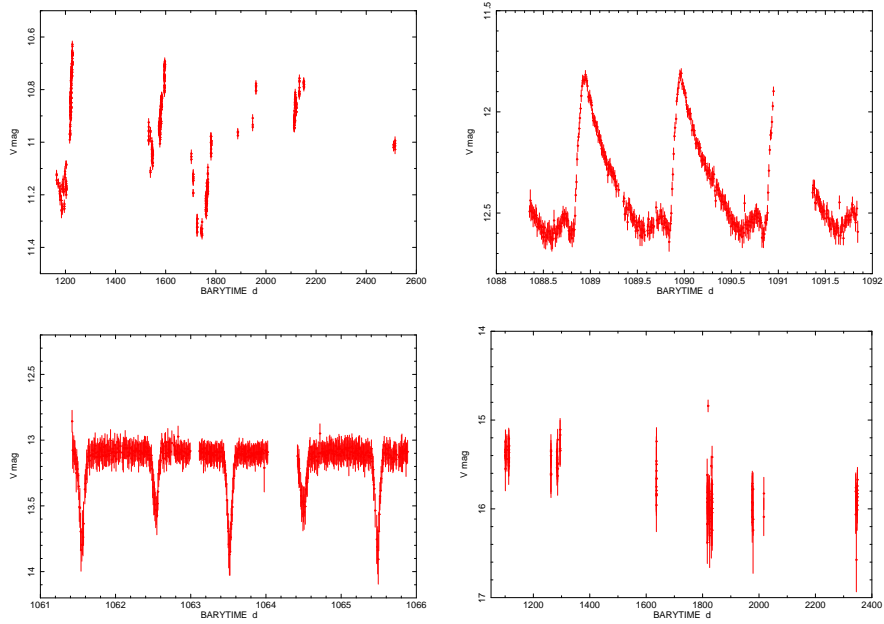

\centering
\includegraphics[width=0.34\textwidth,angle=-90]{obs_post_domingo_1_fig2}
\hfill
\includegraphics[width=0.34\textwidth,angle=-90]{obs_post_domingo_1_fig3}

\vspace{3mm}

\includegraphics[width=0.34\textwidth,angle=-90]{obs_post_domingo_1_fig4}
\hfill
\includegraphics[width=0.34\textwidth,angle=-90]{obs_post_domingo_1_fig5}
\caption{OMC light curves. Top: from left to right, V347 Aql (irregular type) and FY Aqr (pulsating,
         probably RRab). Bottom: from left to right, V809 Cyg (eclipsing binary of Algol type) and
         QSO B1217+0220 (Seyfert 1 galaxy). The origin of the time axis (BARYTIME) is January 1st, 2000.}
\label{fig:curves}
\end{figure}

The photometric accuracy we can achieve in V magnitude with OMC for a typical effective exposure of
300~s ranges from $0.005$ for $V=10$ to $0.04$ for \hbox{$V=14$}. By using a longer effective exposure of
900~s we reach accuracies of $0.026$ for \hbox{$V=14$} and $0.17$ for $V=16$. These values are calculated
in staring mode for isolated sources with a good measurement of the sky background. In dithering
mode (the most usual in INTEGRAL operations), these values are increased by $0.015$~mag, which
corresponds to the accuracy of the flatfield correction.

In Fig.~\ref{fig:curves} we present as an example different types of light curves extracted from our
archive. Note the different ranges in the time axes. For the irregular variable V347~Aql, the
covered range is almost 4 years. In the other extreme, we can see the pulsating star FY~Aqr for
which the total elapsed time covered in the curve is only 4 days.

\begin{acknowledgement}
The activities related to INTEGRAL-OMC are being funded since 1993 by the Spanish National Space
Programme MEC/MICINN (ESP2005-07714-C03-03). This research has made use of the Spanish Virtual
Observatory supported from the Spanish MEC through grants AyA2008-02156, AyA2005-04286.
\end{acknowledgement}

%
%
%

\end{document}